# Growth rates of modern science:

# A bibliometric analysis based on the number of publications and cited references

Lutz Bornmann[1], Rüdiger Mutz[2]

[1] Corresponding author: Division for Science and Innovation Studies, Administrative Headquarters of the Max Planck Society, Munich, Germany, Email: bornmann@gv.mpg.de, Tel: +49 89 2108 1265

[2] Professorship for Social Psychology and Research on Higher Education, ETH Zurich, Zurich, Switzerland, Email: ruediger.mutz@gess.ethz.ch, Tel: +41 44 632 4918

**Abstract**

Many studies in information science have looked at the growth of science. In this study, we re-examine the question of the growth of science. To do this we (i) use current data up to publication year 2012 and (ii) analyse it across all disciplines and also separately for the natural sciences and for the medical and health sciences. Furthermore, the data are analysed with an advanced statistical technique – segmented regression analysis – which can identify specific segments with similar growth rates in the history of science. The study is based on two different sets of bibliometric data: (1) The number of publications held as source items in the Web of Science (WoS, Thomson Reuters) per publication year and (2) the number of cited references in the publications of the source items per cited reference year. We have looked at the rate at which science has grown since the mid-1600s. In our analysis of cited references we identified three growth phases in the development of science, which each led to growth rates tripling in comparison with the previous phase: from less than 1% up to the middle of the 18th century, to 2 to 3% up to the period between the two world wars and 8 to 9% to 2012.

# 1 Introduction

Many studies in information science have looked at the growth of science (Evans, 2013). Tabah (1999) offers an overview of the literature which groups these studies under the label "the study of literature dynamics" (p. 249): "The information science approach is to follow the published literature and infer from the growth of the literature the movement of ideas and associations between scientists" (Tabah, 1999, p. 249). Price (1965; 1951, 1961) can undoubtedly be seen as a pioneering researcher on literature dynamics (de Bellis, 2009). Price analysed the references listed in the 1961 edition of the Science Citation Index (SCI, Thomson Reuters) and the papers collected in the *Philosophical Transactions of the Royal Society of London*. His results show that science is growing exponentially (in a certain period by a certain percentage rate) and doubles in size every 10 to 15 years. The exponential growth in science established by Price has become today a generally accepted thesis which has also been confirmed by other studies (Tabah, 1999).

In this study, we want to re-examine the question of the growth of science. To do this we will (i) use current data up to publication year 2012 and (ii) analyse it across all disciplines and also separately for the natural sciences and for the medical and health sciences. Furthermore, the data will be analysed with an advanced statistical technique – segmented regression analysis – which can identify specific segments with similar growth rates in the history of science. The study is based on two different sets of data: (1) The number of publications held as source items in the Web of Science (WoS, Thomson Reuters) per publication year and (2) the number of cited references in the publications of the source items per cited reference year (Bornmann & Marx, 2013; Marx, Bornmann, Barth, & Leydesdorff, in press). The advantage of using cited references rather than source items is that they can give insight into the early period of modern science. There is no database available which covers publications (source items) from the early period. The disadvantage of using cited references is that the literature which has not been cited yet is not considered. Furthermore, publishing in the early period is inferred by todays citing (here: in the period from 1980 to 2012).

## 2  Methods

Publications are very suitable source of data with which to investigate the growth rates of science: "Communication in science is realized through publications. Thus, scientific explanations, and in general scientific knowledge, are contained in written documents constituting scientific literature" (Riviera, 2013, p. 1446). Having a paper published in a journal is an integral part of being a scientist: "[It] is a permanent record of what has been discovered, when and by which scientists – like a court register for science – [and it] shows the quality of the scientist's work: other experts have rated it as valid, significant and original" (Sense About Science, 2005). Because "efficient research requires awareness of all prior research and technology that could impact the research topic of interest, and builds upon these past advances to create discovery and new advances" (Kostoff & Shlesinger, 2005, p. 199), cited references in the publications are also an important source of data with which to examine scientific growth. An increase in the number of cited references indicates that there are more citing and/or cited publications.

Our study is based on all the publications from 1980 to 2012 and the cited references in these publications. The data is taken from an in-house database belonging to the Max Planck Society (Munich, Germany) based on WoS. It was established and is maintained by the Max Planck Digital Library (MPDL, Munich, Germany). As the data prepared by the MPDL relates to publications (and their cited references) since 1980, it was only possible to include these publications (and their cited references) in the analysis. The first step in the study was to select all the publications (all document types) that appeared between 1980 and 2012 (38,508,986 publications) and determine the number of publications per year. The second step was to select the cited references in the publications from 1980 to 2012 and to determine the number of cited references per year (from 1650 to 2012) (755,607,107 cited references in total). The annual number of publications or cited references formed the basis for the (segmented) regression analyses (van Raan, 2000) – the third step in the analysis.

Based on the annual number of publication, a growth model $y(t)=b_0*exp(b_1*(t-1980))$ was estimated by a nonlinear regression using SAS PROC NLIN (SAS Institute Inc., 2011), where the intercept $b_0$ equals $y(0)$ – the outcome in the year 1980. The model converged: overall 96% of the total variance of the annual number of publications could be explained by the regression model.

Segmented regression analysis was used to determine different segments of growth development in the cited references within the annual time series (Bornmann, Mutz, & Daniel, 2010; Brusilovskiy, 2004; Lerman, 1980; McGee & Carleton, 1970; Mutz, Guilley, Sauter, & Nepveu, 2004; Sauter, Mutz, & Munro, 1999; Shuai, Zhou, & Yost, 2003). In the model estimations, the logarithmised number of cited references per year forms the dependent variable. In mathematical and statistical terms we assumed a simple exponential growth model which considers separate segments in the time series (e.g., a segment with a decline around both World Wars, WW). This model can be formulated as a differential equation $f`(t)=b_1*f(t)$ where $b_1$ is the growth constant or the multiplication factor and t is the time (cited reference year). The change $f(t)$ in a period $t_1-t_0$ is therefore proportional to the status at the starting point in time $t_0$. The solution of the differential equation is an exponential function: $y(t)=y(0)*exp(b_1*t)$. The growth rate in percent $(y(t)-y(0))/y(0)$ is $exp(b_1)-1$. The doubling time is the amount of time required for an outcome to double in size ($t=ln(2)/b_1$). Logarithmising the function $y(t)$ results in a linear function $ln(y)=b_0+b_1*t$ with $b_0=log(y(0))$, the parameters of which can be estimated with a linear regression.

In the segmented regression different segments can be identified with different regression coefficients, where both the breakpoints (cited reference year) a of the segments as well as the growth constant $b_1$ of each segment is estimated. For example, let log_y the logarithmic transformed annual number of cited references, 'year' the cited reference year, and $a_1/a_2$ the breakpoints of differentiating three segments. Then, we need to estimate the unknown regression parameters $b_0$, $b_1$, $b_2$ and $b_3$, and the breakpoints $a_1$ and $a_2$ by minimizing the following objective function – in particular the sum of squared residuals (Brusilovskiy, 2004, p. 2):

IF year<=$a_1$ THEN log_y=$b_0$+$b_1$*year+$\varepsilon$;  (1)

ELSE IF year <=$a_2$ THEN log_y= $b_0$+$b_1$*$a_1$+$b_2$*(year-$a_1$)+ $\varepsilon$;

ELSE IF year <=$a_3$ THEN log_y=$b_0$ + $b_1$*$a_1$ + $b_2$*($a_2$-$a_1$)+$b_3$*(year-$a_2$)+ $\varepsilon$

$\varepsilon$ ~ i.i.d. N(**0**, **I**$\sigma^2$)

In the function the residuals $\varepsilon$ are assumed to be multivariate normally distributed with a zero mean vector and a covariance matrix with identical variances $\sigma^2$ (homoscedasticity) and zero covariance (no autocorrelations of the residuals) overall and across the segments. These are rather strong assumptions for time series data. Given the high proportion of explained variance (low proportion of residual variance in this study) and low standard errors, which we generally observed in the regression analyses, the violations of the assumptions were not considered with the risk of slightly higher standard errors. The model parameters were estimated by the least squares method (Gauss-Newton) under the restriction that the breakpoints $a_1$, $a_2$, … are ordered (e.g., 1690<$a_1$<$a_2$<2012). In order to avoid local minima of the estimation procedure a grid of different starting values for the parameters was used. The regression constant $b_0$ was erased in our analysis to enhance the fit of the model.

The following criteria were used to choose the number of segments: First, by visually inspecting the data it was obvious that the relationships are strongly linear with different slopes within three or four segments (see, e.g., Figure 2). Second, statistical criteria were used. In the case of least-squares estimations the amount of explained variance ($R^2$) is the method of choice. In this study, models were selected, where the amount of explained variance no longer substantially increased. In other words, segmented regression models with three breakpoints could explain 99% of the total variance of the annual number of references. Thus, it was possible to divide the series of the annual number of cited references in the time interval from 1650 to 2012 into four segments.

Differences between disciplines (natural sciences and medical and health sciences) (D) regarding the slope parameters are tested by expanding equation 1 by corresponding interaction terms with disciplines (e.g., $b_1*D*year$) for each slope parameter ($b_1$, $b_2$, $b_3$, …). The disciplines were dummy-coded (natural sciences: D=0, medical and health sciences: D=1).

The statistical analyses were performed using the SAS procedure PROC NLIN (SAS Institute Inc., 2011).

## 3 Results

Figure 1 shows the exponential growth of global scientific publication output for the period 1980 to 2012 (33 years). While the dots represent the observed values, the predicted values are the result of the regression analysis. As the result of the segmented regression analysis (Table 1) shows, the global scientific publication output is growing at a rate of approximately 3% annually. The volume of publications doubles approximately every 24 years. A similar result – exponential growth of publications – is also reported by Pautasso (2012) for 18 biological sub-fields.

However with the model in Table 1 we are looking at a relatively short and recent period of scientific activity in society. The analysis of cited references from the publications allows us to start the analysis of scientific activity much earlier. According to van Raan (2000) scientific evolution started "after the Scientific Revolution in 16th century Europe" (p.

347). However, van Raan (2000) starts the analysis of cited references considerably later, in 1800. In this study, we start with the analysis of cited references in the middle of the 17th century. This period saw the development of the institutionalised structures of modern science with publication of the results of scientific work in journals and manuscripts undergoing critical peer review before publication (Bornmann, 2011). In peer review, active scientists in the same discipline review the scientific work of each other and decide on success and failure. The use of the peer review procedure since the middle of the 17th century is an important step towards an independent science system from other areas of society, for example from church (Gould, 2013). According to Popper (1988) we can assume that it is rational criticism – institutionalized as peer review procedures – driven by the idea of truth that characterises modern science. Gattei (2009) says that the "growth of knowledge and criticism are closely interconnected" (p. 2).

Figure 2 shows the segmented growth of the annual number of cited references between 1650 and 2012. The observed values and those predicted by the model are visualised. As Table 2 shows, regression analysis has identified four different segments with different growth rates: The first phase from 1650 to approximately 1750 is characterised by a relatively low growth rate of less than 1%, with cited references doubling in 150 years approximately. It is followed by a phase with a growth rate of slightly less than 3% which lasts until the period between WWI and WWII. Interestingly, these two segments identified by the regression analysis correspond to phases of economic growth and the start of the Industrial Revolution in Europe in about 1750. "Pre-1750 growth was primarily based on Smithian and Northian effects: gains from trade and more efficient allocations due to institutional changes. The Industrial Revolution, then, can be regarded not as the beginnings of growth altogether but as the time at which technology began to assume an ever-increasing weight in the generation of growth and when economic growth accelerated dramatically. An average growth rate of 0.15–0.20% per annum, with high year-to-year variation and frequent setbacks was replaced by a much more steady growth rate of 1.5% per annum or better" (Mokyr, 2005, p. 1118).

The second phase in Figure 2 is followed by a third phase with the strongest growth rate of approximately 8% and which lasts to the start of the 21st century. Doubling time for cited references in this phase is approximately 9 years. The final phase has a negative growth rate. Around 1980, the curve in Figure 2 starts to flatten off and then fall significantly: from 1980, more publications are included in the regression analysis each year, which in many cases refer to publications which could not yet be cited in the previous year. The fourth segment therefore reflects "time-dependent database characteristics in, for instance, the registration of references" (van Raan, 2000, p. 348).

To establish in how far the use of all the publications since 1980 has an influence on the growth rate of science (particularly in the last 30 years), we have extracted the annual growth rates of the cited references given in the publications for 2012 in Figure 3. 1,859,648 publications (from 2012) and 53,345,550 cited references between 1650 and 2012 were included in this analysis. As illustrated by a comparison of the results in Table 3 with the results in Table 2 the regression analysis identified four segments with different growth rates even though only one year was taken into account. While the growth rates for the medium and younger segment hardly differ, (we are looking here at just the segments with a positive growth rate), when only one year is taken into account, the growth rate is much lower in the oldest segment than in the analysis of citing publications from 1980 to 2012. By taking just one year into account, the number of cited references particularly in this segment falls as generally speaking, very old publications are rarely cited (Marx, 2011). In contrast to Figure 2 we see in Figure 3 an abrupt decline in the number of cited references from around 2010. van Raan (2000), who observes a similar fall a few years before 1998 in his data, explains this drastic decline thus: "It is well known, particularly in the natural science and medical fields, that publications of a given year (here: 1998) have a peak in the distribution of their references around the age of three years. Therefore, [in the figure] in the last three years the number of references will decrease" (p. 350).

According to van Raan (2000) the growth rates in Figure 2 and Figure 3 reflect two processes: "*ageing* (scientists will be increasingly less interested in increasingly older literature), and *growth* (there were simply much less papers published in 1930 as there are in 1990, so there is less to be cited to earlier years) " (p. 351). Tabah (1999) also discusses these two processes in connection with growth in science; they influence each other as described here: "Specifically, the faster a literature or even a given journal grows (in terms of number of articles published per year), the more rapidly it ages" (p. 250, see also Egghe & Rousseau, 2000). Similarly van Raan (2000) writes: "Although undoubtedly ageing of earlier published work is part of reality, another part of reality is that in earlier times there were many fewer documents published than in recent times. Thus, the time-dependent distribution of references will always be a specific combination of the ageing and growth phenomena of science, or better said: of scientific literature" (pp. 347-348).

In earlier analyses of the growth rate of science, the time before, between and WWI and WWII has been of particular interest (Larsen & von Ins, 2010; Price, 1965; van Raan, 2000). In conformity with our results, other studies have also shown a significant fall in the number of cited references during this period. Furthermore, this and other studies, as van Raan (2000) writes, also show that the two dips are "immediately followed by an astonishing 'recovery' of science after the wars toward a level more or less extrapolated from the period before the wars" (p. 347). However the results for the period between WWI and WWII, from around 1920 to 1935, revealed some inconsistencies: "In Price's analysis this period shows a 'hill' when extrapolating the exponential increase of the pre-World War I period. This 'hill' has almost disappeared in our analysis" (van Raan, 2000, p. 349). Like Price's, our study shows that a significant increase in the number of cited references in the period between the wars has several consequences, one of which is that the third segment with a growth rate of around 8% (see Table 2) does not begin in the 'Big Science' period (see above), but a few years earlier. According to our data, WWII is only a temporary low point in the significant growth of science since the period between the wars.

In our final analysis we looked at the extent to which we could determine whether there were differences between growth rates in different disciplines. The main categories of the Organisation for Economic Co-operation and Development (2007) (OECD) were used as a subject area scheme for this study. A concordance table between the OECD categories and the WoS subject categories is provided by InCites (Thomson Reuters) (see http://incites-help.isiknowledge.com/live/globalComparisonsGroup/globalComparisons/subjAreaSchemes Group/oecd.html). The OECD scheme enables the use of six broad subject categories for WoS data: (1) natural sciences, (2) engineering and technology, (3) medical and health sciences, (4) agricultural sciences, (5) social sciences, and (6) humanities. However, neither the numbers for engineering and technology nor those for the social sciences and humanities were included in this study. According to the Council of Canadian Academies (2012), the usefulness of citation impact indicators depends on the extent to which the research outputs are covered in bibliometric databases, and this coverage varies by subject category. The coverage tends to be high in the natural and life sciences, which place a high priority on journal publications. In engineering, the social sciences and humanities, where the publication of books, book chapters, monographs, conference proceedings etc. is more traditional, the extent of the coverage is reduced (Lariviere, Gingras, & Archambault, 2006).

We therefore undertook discipline-specific analyses for the natural sciences and the medical and health sciences. 15,435,641 publications from 1980 to 2012 and corresponding 379,294,777 cited references from 1650 to 2012 were included in the analyses for the natural sciences. The corresponding figures for the medical and health sciences were 12,796,558 publications and 256,164,353 cited references. The results of the analyses are shown in Figure 4, Figure 5, Table 4, and Table 5. The results are very similar. The regression analysis identified 3 segments with interpretable content up to the beginning of the 21st century, with similar start and end points. Only the middle segment, which exhibits a higher level of scientific activity compared to the oldest segment, begins slightly earlier in the natural sciences than in the medical and health sciences (in 1720 rather than 1750). Furthermore, the growth rates in the medical and health sciences since the 18th century have been minimally higher than in the natural sciences. The doubling times in the medical and health sciences are correspondingly slightly lower than in the natural sciences.

The results of the regression analysis in which the interaction terms with disciplines are included confirm the similarities. The interaction effects (results are not shown here) disciplines × publication years for each segment are statistically significant, but they are too small to justify any statements about real differences between the disciplines. The statistical significance results (more or less) from the high proportion of explained variance and the associated very low standard errors of the parameters.

## 4 Discussion

In this study we have looked at the rate at which science has grown in terms of number of publications and cited references since the mid-1600s. In our analysis we identified three growth phases in the development of science, which each led to growth rates tripling in comparison with the previous phase: from less than 1% up to the middle of the 18th century, to 2 to 3% up to the period between the two world wars and 8 to 9% to 2012. As the growth rate of the cited references in the third segment is significantly higher than that based on the source segments in this study (approximately 3%), we can assume that WoS only covered a small part of the total publications. We can further suppose that the number of early publications is underestimated and the publications' growth rate is overestimated in the cited reference analysis because of *aging* (see above). van Raan (2000) reports, based on analyses of cited references, a growth rate of around 10% for the period from 1800 to the mid-1990s.

We can only confirm this result for the last 70 years (and not for the last 200 years). However, when considering the studies by Price (1965) and by van Raan (2000) and our study, it should be noted that scientific growth since the beginning of modern science has not been measured directly with the publication volume, but indirectly with the number of cited references. Unfortunately, there is currently no literature database containing every publication since the beginning of modern science to today and which can be used for statistical analysis.

Our results support estimates such as those made in a comment piece by Frazzetto (2004) about the far-reaching changes that have taken place in science over the last century:

> "Without doubt, science has put on a new face in the past century. It has come to occupy a central role in society and now enjoys a privileged position among the knowledge-producing disciplines. As a consequence, the resources devoted to science have increased hugely. There has also been, in general, a visible shift in the way in which science is organized and in how it produces knowledge. Science has become 'big', global and complex … and the emphasis has shifted from the individual scientist to collaborative work. Similarly, scientific knowledge production has become more costly, and forms of funding have subsequently changed, with industry and venture capital now providing the bulk of financial resources" (Frazzetto, 2004, p. 19).

An end to this development cannot be foreseen (with new highly-productive players such as China and India moving swiftly into the science community).

Finally, we would like to mention three limitations of our study:

The first limitation concerns the use of publications to measure scientific growth. There are advantages and disadvantages to using this data, as described by Tabah (1999): "Although counting publications is simple and relatively straightforward, interpretation of the data can create difficulties that have in the past led to severe criticisms of bibliometric methodology … The main problems concern the least publishable unit (LPU), disciplinary variance, variance in quality of work, and variance in journal quality" (p. 264). Publications are not singular entities that have static forms. Publication practices differ, not only between disciplines, but also within disciplines. Publication numbers and both publishing styles as well as citing styles have developed through history (de Bellis, 2009). Add to this the explosion in publishing numbers linked to external factors in the past decades, related to the "Audit Society" and "New Public Management", with "salami sliced publishing", and the notion of "least publishable unit" (Bornmann & Daniel, 2007). Not to mention that at least during the first 100 years of publishing in *Philosophical Transactions*, it was not the scientists that published their work, but the editor, that received letters of talks read to the community that were published. There is of course the option of using other data than publications to measure scientific growth such as the number of scientists. However, as a rule this other data is not more suitable than bibliometric data as it has its own limitations (for example, there is no database that can provide reliable information about the number of scientists practising modern science since its beginnings to this day).

The second limitation refers to the view of "growth" as an "increase in numbers" (publications, cited references, or scientists). This study views science through an internal perspective, where all differences are expected to be the result of internal aspects of a static research practice in "modern science", as opposed to a mixture of internal (analytic) practices and external (sociological, historical, psychological) practices that continuously have altered the ways of viewing science in the same four centuries that this study focuses on. Furthermore, it is not clear whether an "increase in numbers" is directly related to an "increase of actionable knowledge", for example for reducing needs, extending our knowledge about nature in some lasting way or some other "higher purposes" (Bornmann, 2012, 2013).

The third limitation concerns the database used here: WoS. According to van Raan (2000) "characteristics of the instruments (in this case: the database) will also be reflected in the measurements. Even worse, characteristics of databases may change in the course of time"

(p. 348). The WoS (or the SCI) has been in operation for several decades and has changed over the years. For example, the coverage of journals has been extended. The growth of science is not the only explanation for an increasing number of publications in the data, as Michels and Schmoch (2012) note: "By subdividing the journals into different categories, it was possible to distinguish which increases are related to a growth of science and which to the database provider's policy of achieving a broader coverage of journals. The number of articles in the observation period [2000-2008] grew by 34 percent in total, a remarkable increase for this short period. If the number of articles in the categories 'start' and 'new', which represent the growth of science, are added together, the growth rate in this period is 17 percent" (p. 841). As this study is mainly based on an analysis of cited references, we are assuming that the changes of the database characteristics do not exert a major influence on the results. As we have shown, even substantial changes to the data, such as limiting it to one year or one field in the citing papers, does not have a significant influence on the results arrived at with the cited references.

# Acknowledgements

The data used in this paper is from a bibliometrics database developed and maintained by the Max Planck Digital Library (MPDL, Munich) and derived from the Science Citation Index Expanded (SCI-E), Social Sciences Citation Index (SSCI), Arts and Humanities Citation Index (AHCI) prepared by Thomson Reuters (Scientific) Inc. (TR®), Philadelphia, Pennsylvania, USA: ©Copyright Thomson Reuters (Scientific) 2014. The bibliometric data for the study has been downloaded on February 2014. We would like to thank two anonymous reviewers for their valuable feedback to improve the paper. We received extensive and very interesting comments from both reviewers. Some comments were directly incorporated into the manuscript.

Table 1. Estimation of the exponential growth model (time interval for publications: from 1980 to 2012)

| Parameter | Estimate | SE | 95% confidence interval | % growth rate | Doubling time [year] |
|---|---|---|---|---|---|
| $b_0$ | 702,880* | 17,430.6 | 667,330 – 738,430 | | |
| $b_1$ | 0.029* | 0.001 | 0.027 - 0.031 | 2.96% | 23.7 |

Notes. $R^2$=.96
*p<.05

Table 2. Estimation of the segmented regression (time interval for citing publications: from 1980 to 2012; time interval for cited references: from 1650 to 2012)

| Parameter | Estimate | SE | 95% confidence interval | % growth rate | Doubling time [year] |
|---|---|---|---|---|---|
| $a_1$ | 1753.3* | 2.34 | 1748.7 - 1757.9 | | |
| $a_2$ | 1926.1* | 1.22 | 1923.7 - 1928.5 | | |
| $a_3$ | 2000.6* | 0.50 | 1990.6 - 2001.6 | | |
| $b_1$ | 0.005* | 0.000 | 0.004 - 0.005 | 0.45% | 155.8 |
| $b_2$ | 0.023* | 0.000 | 0.023 - 0.024 | 2.35% | 29.9 |
| $b_3$ | 0.078* | 0.001 | 0.076 - 0.081 | 8.13% | 8.9 |
| $b_4$ | -0.22* | 0.02 | -0.258 - -0.179 | -19.62% | - |

Notes. $R^2$=.99
*p<.05

Table 3. Estimation of the segmented regression (time interval for citing publications: 2012; time interval for cited references: from 1650 to 2012)

| Parameter | Estimate | SE | 95% confidence interval | % growth rate | Doubling time [year] |
|---|---|---|---|---|---|
| $a_1$ | 1739.2* | 1.90 | 1735.5 – 1742.9 | | |
| $a_2$ | 1942.2* | 0.86 | 1940.5 - 1943.9 | | |
| $a_3$ | 2010.9* | 0.16 | 2010.6 - 2011.2 | | |
| $b_1$ | 0.003* | 0.000 | 0.003 - 0.003 | 0.27% | 253.9 |
| $b_2$ | 0.022* | 0.000 | 0.021 - 0.022 | 2.19% | 31.9 |
| $b_3$ | 0.088* | 0.001 | 0.086 - 0.090 | 9.20% | 7.9 |
| $b_4$ | -1.310* | 0.290 | -1.874 - -0.745 | -73.01% | - |

Notes. $R^2$=.99
*p<.05

Table 4. Estimation of the segmented regression for the natural sciences (time interval for citing publications: from 1980 to 2012; time interval for cited references: from 1650 to 2012)

| Parameter | Estimate | SE | 95% confidence interval | % growth rate | Doubling time [year] |
|---|---|---|---|---|---|
| $a_1$ | 1718.3* | 2.39 | 1713.6 - 1723.0 | | |
| $a_2$ | 1925.5* | 1.77 | 1922.0 - 1929.0 | | |
| $a_3$ | 1999.2* | 0.75 | 1998.1 - 2001.1 | | |
| $b_1$ | 0.003* | 0.000 | 0.003 - 0.003 | 0.29% | 231.1 |
| $b_2$ | 0.029* | 0.000 | 0.028 - 0.030 | 2.94% | 23.9 |
| $b_3$ | 0.081* | 0.002 | 0.077 - 0.084 | 8.37% | 8.7 |
| $b_4$ | -0.189* | 0.025 | -0.238 - -0.139 | -14.20% | - |

Notes. $R^2$=.99
*p<.05

Table 5. Estimation of the segmented regression for medical and health sciences (time interval for citing publications: from 1980 to 2012; time interval for cited references: from 1650 to 2012)

| Parameter | Estimate | SE | 95% confidence interval | % growth rate | Doubling time [year] |
|---|---|---|---|---|---|
| $a_1$ | 1753.5* | 2.23 | 1749.1 – 1757.8 | | |
| $a_2$ | 1928.6* | 1.60 | 1925.4 - 1931.7 | | |
| $a_3$ | 2002.1* | 0.63 | 2000.9 - 2003.4 | | |
| $b_1$ | 0.003* | 0.000 | 0.003 - 0.003 | 0.28% | 250.2 |
| $b_2$ | 0.031* | 0.000 | 0.030 - 0.032 | 3.10% | 22.3 |
| $b_3$ | 0.089* | 0.002 | 0.086 - 0.093 | 9.35% | 7.8 |
| $b_4$ | -0.297* | 0.038 | -0.37 - -0.222 | -25.67% | - |

Notes. $R^2=.99$
*$p<.05$

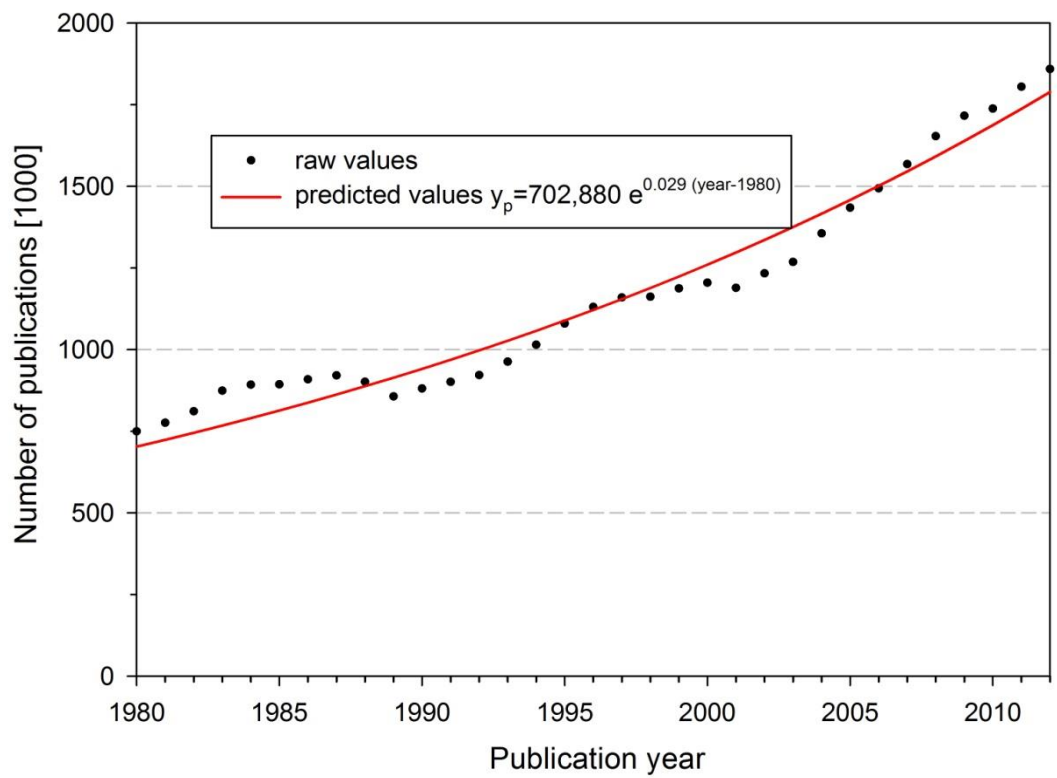

Figure 1. Exponential growth of scientific output from 1980 to 2012

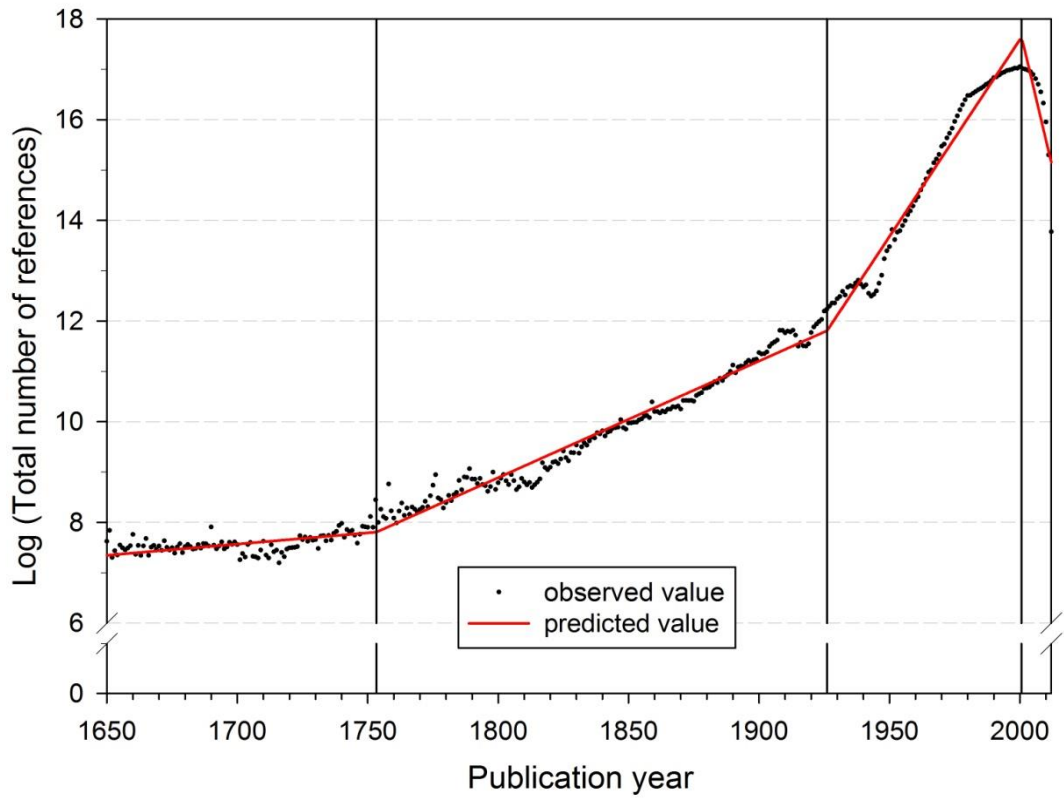

Figure 2. Segmented growth of the annual number of cited references from 1650 to 2012 (citing publications from 1980 to 2012)

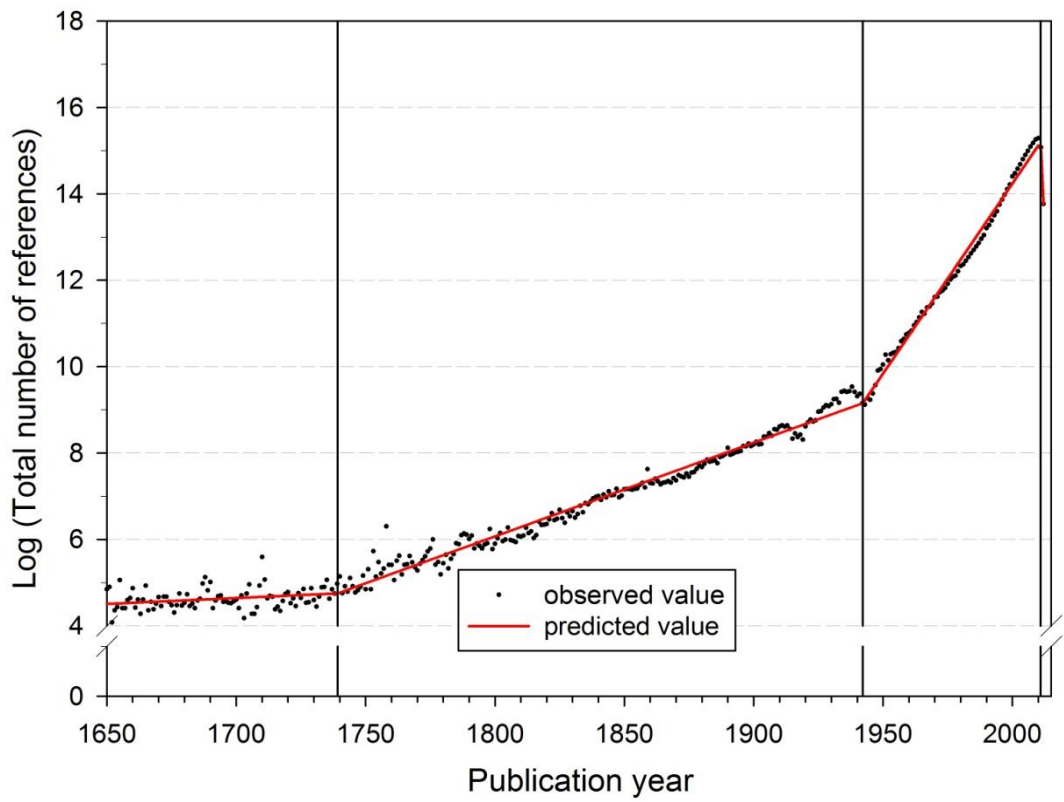

Figure 3. Segmented growth of the annual number of cited references from 1650 to 2012 (citing publications from 2012)

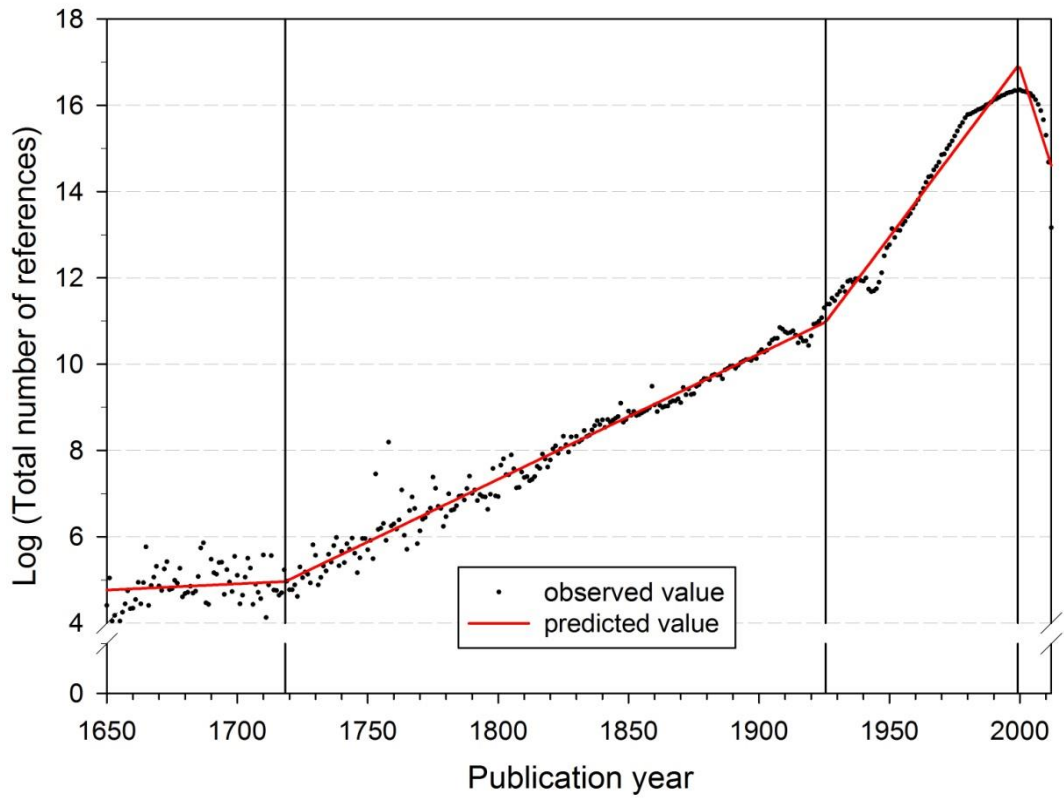

Figure 4. Segmented growth of the annual number of cited references from 1650 to 2012 in the natural sciences (citing publications from 1980 to 2012)

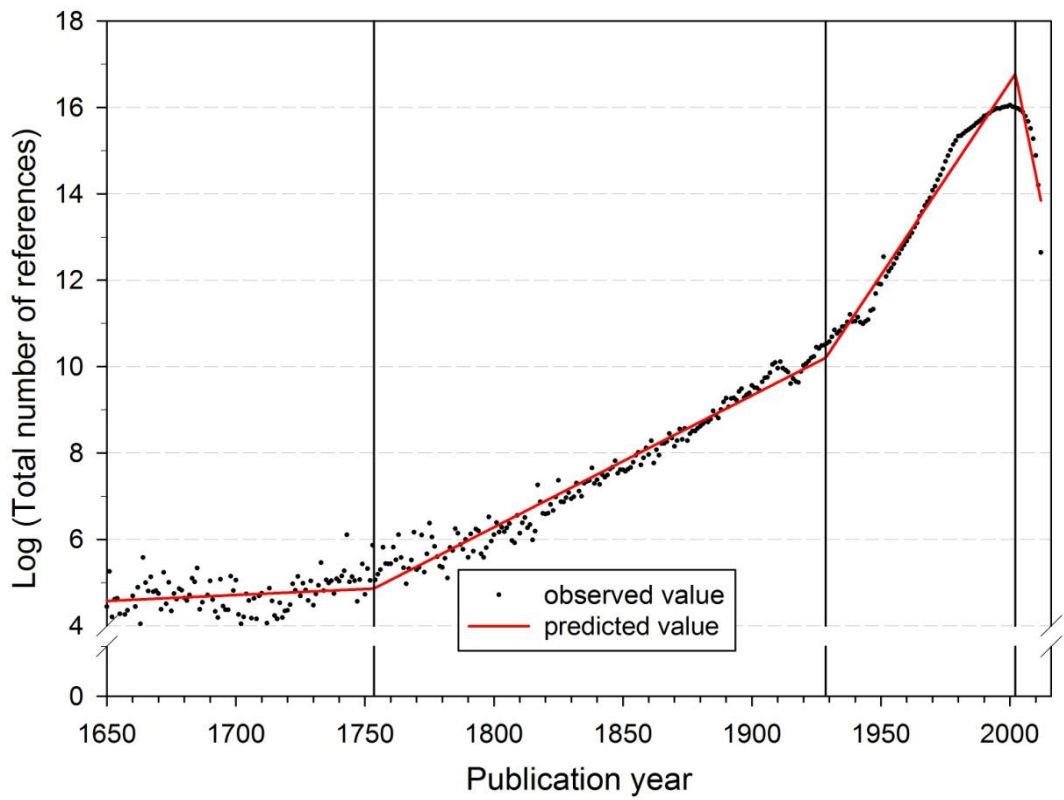

Figure 5. Segmented growth of the annual number of cited references from 1650 to 2012 in the medical and health sciences (citing publications from 1980 to 2012)